\documentclass[aps,prl,reprint,floatfix,citeautoscript,longbibliography,superscriptaddress]{revtex4-1}
\usepackage{graphicx}
\usepackage{bm,amsmath,amssymb,mathrsfs,dcolumn}
\usepackage{gensymb}
\usepackage[usenames]{color}
\usepackage{subfigure}
\usepackage{ulem}
\usepackage{soul,xcolor}
\usepackage{hyperref}
\hypersetup{colorlinks=true, linkcolor=blue, citecolor=red, urlcolor=blue}
\setstcolor{red}

\begin{document}

\title{Enhancement of antiferromagnetic magnon-magnon entanglement by cavity cooling}
\author{H. Y. Yuan}
\email[Electronic address: ]{yuanhy@sustech.edu.cn}
\affiliation{Department of Physics, Southern University of Science and Technology, Shenzhen 518055, China}
\author{Shasha Zheng}
\affiliation{State Key Laboratory for Mesoscopic Physics, School of Physics $\&$ Collaborative Innovation Center of Quantum Matter, Peking University, Beijing 100871, China}
\affiliation{Beijing Academy of Quantum Information Sciences, Haidian District, Beijing 100193, China}
\affiliation{Collaborative Innovation Center of Extreme Optics, Shanxi University, Taiyuan, Shanxi 030006, China}
\author{Zbigniew Ficek}
\affiliation{Quantum Optics and Engineering Division, Institute of Physics, University of Zielona G\'{o}ra, Zielona G\'{o}ra, Poland}
\author{Q. Y. He}
\email[Electronic address: ]{qiongyihe@pku.edu.cn}
\affiliation{State Key Laboratory for Mesoscopic Physics, School of Physics $\&$ Collaborative Innovation Center of Quantum Matter, Peking University, Beijing 100871, China}
\affiliation{Beijing Academy of Quantum Information Sciences, Haidian District, Beijing 100193, China}
\affiliation{Collaborative Innovation Center of Extreme Optics, Shanxi University, Taiyuan, Shanxi 030006, China}
\author{Man-Hong Yung}
\email[Electronic address: ]{yung@sustech.edu.cn}
\affiliation{Institute for Quantum Science and Engineering and Department of Physics, Southern University of Science and Technology, Shenzhen, 518055, China}
\affiliation{Shenzhen Key Laboratory of Quantum Science and Engineering, Shenzhen 518055, China}
\date{\today}

\begin{abstract}
Magnon-photon coupling has been experimentally realized inside a cavity and the emerging field known as cavity spintronics has attracted significant attention for its potential docking with quantum information science. However, one seldom knows whether this coupling implies an entanglement state among magnons and photons or not, which is crucial for its usage in quantum information. Here we study the entanglement properties among magnons and photons in an antiferromagnet-light system and find that the entanglement between magnon and photon is nearly zero while the magnon-magnon entanglement is very strong and it can be even further enhanced through the coupling with the cavity photons. The maximum enhancement occurs when the antiferromagnet reaches resonant with the light. The essential physics can be well understood within the picture of cavity induced cooling of magnon-magnon state near its joint vacuum with stronger entanglement. Our results are significant to extend the cavity spintronics to quantum manipulation and further provide an alternate to manipulate the deep strong correlations of continuous variable entanglement with a generic stable condition and easy tunability.
\end{abstract}

\maketitle

{\it Introduction.---} Antiferromagnetic (AFM) spintronics emerges for its better stability and fast dynamics over its ferromagnetic counterpart~\cite{Jungwirth2016, Baltz2018}. Of particular interest are antiferromagnetic spin waves (magnons) that show much richer physics than ferromagnets, such as the spin pumping at the interface of an AFM/normal metal bilayer~\cite{RCheng2014}, magnon spin current enhancement through an AFM layer~\cite{Wang2014, Hahn2014, Lin2016, Frangou2016}, long-distance magnon transport~\cite{Takei2014,Hanwei2018,Klaui2018}, and magnon-driven magnetic structure motion~\cite{Tveten2014,Yu2018,Yuan2019}. It has recently been theoretically proposed~\cite{Yuan2017apl} and experimentally verified~\cite{Merg2017} that antiferromagnetic magnons can be strongly coupled to the light by placing an AFM element into a cavity, which serves a promising candidate to realize coherent information transfer between magnons and photons for its superb stability and tunability of magnons properties through external knobs. One promising application of the hybridized magnon-photon polariton is to connect it with the quantum information science~\cite{Yao2017,Wang2018}, similar to the cavity electrodynamics that rose 20 years ago and has found its role in the implementation of quantum qubits and quantum computing circuits~\cite{Gu2017}. To push the development of cavity spintronics along this line, it is crucial to have a clear insight into the quantum correlations among magnons and photons, especially their entanglement properties, which are central resources for quantum computing and quantum technologies.

Intuitively, one may think that the quantum coupling between magnons and photons could always result in quantum entanglement between them, however, this is not essentially true. Here both magnons and photons are bosons with continuous excitation spectrum while determining the existence and quantity of entanglement between these continuous modes are highly non-trivial issues~\cite{Duan2000,Simon2000,Hillery2006,Adesso2007,He2012,Sete2012,Li2018}. As widely investigated in quantum optics~\cite{He2012,Adesso2007}, if the two bosonic modes ($a$ and $b$) are coupled in a beam-splitter-type ($a^\dagger b + a b^\dagger$), there is no entanglement between them in a steady state, while if their interaction is parametric-down-conversion-type ($a^\dagger b^\dagger + a b$), the modes can be entangled. In an AFM, the situation is more complex since the two types of magnons are mutually coupled to each other and are also coupled, but in different way, to the photon mode. Therefore, it is desirable to determine how magnons and photons interplay to entangle with each other in this three-mode system.

In this letter, we study the entanglement properties among magnons and photons in an AFM-light system and find that the magnon-photon entanglement is very weak near the anticrossing point of the spectrum. Surprisingly, the magnon-magnon entanglement is very strong at this point and becomes even stronger than the magnon-magnon entanglement in the absence of light, which is ascertained to be a unique feature of the deep strong coupling between magnons. The essential physics can be well understood using the cavity cooling picture, where cavity photons effectively cool the magnons near their vacuum state exhibiting a much stronger entanglement. These results may be significant if one tends to extend the cavity spintronics to quantum manipulation, where the entanglement among magnons is an important resource.

\begin{figure}
  \centering
  \includegraphics[width=0.4\textwidth]{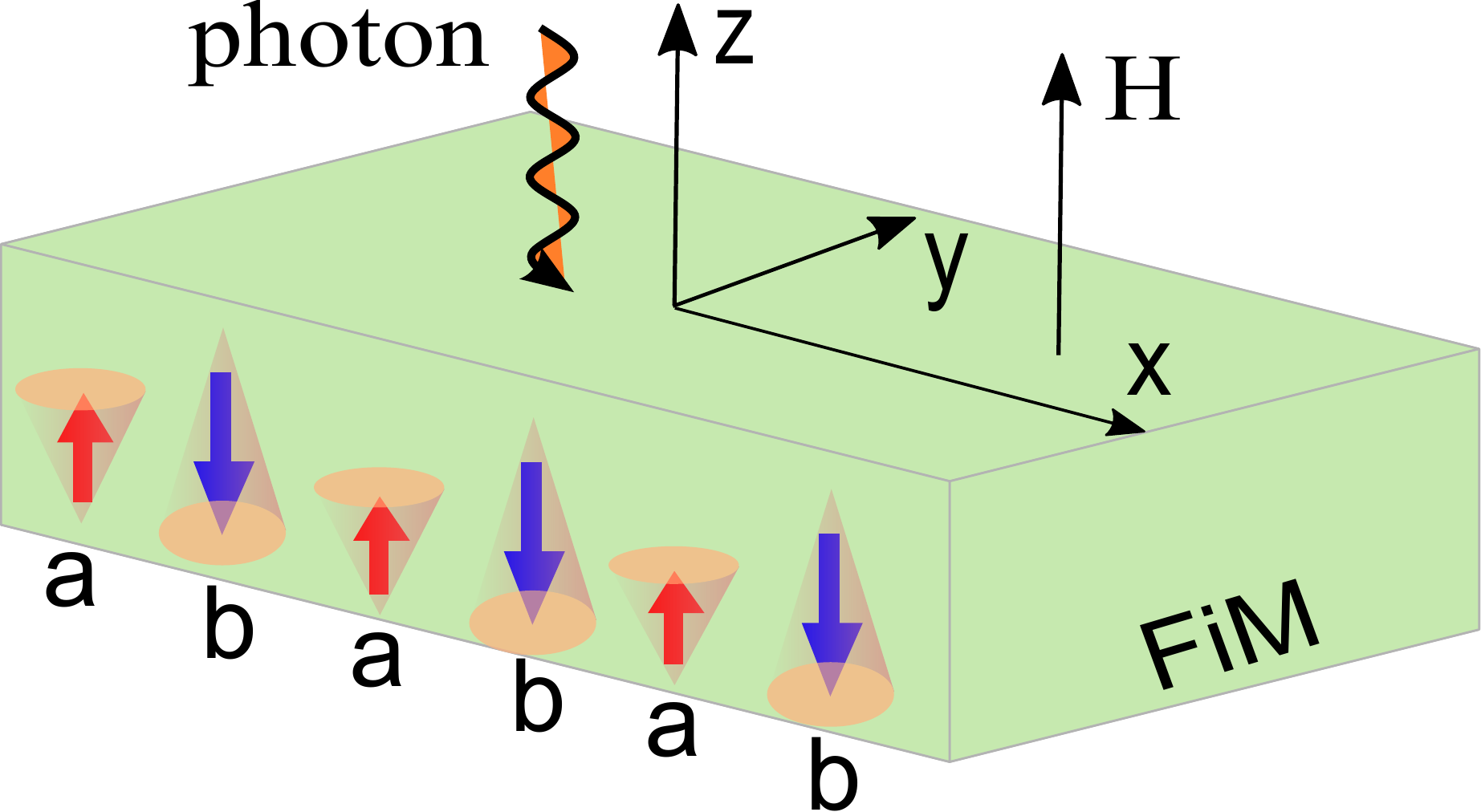}
  \caption{Scheme of a two-sublattice ferrimagnet and its coupling with an electromagnetic wave. The magnetic moments on the two sublattices (red and blue arrows) pointing along $\pm z$ directions represent the classical ground state of the system.}
\label{fig1}
\end{figure}

{\it General formalism.---} We begin with presenting a general theory of the magnet-light interaction in a two-sublattice magnet as shown in Fig.~\ref{fig1}. Then, we concentrate on the AFM case followed by a short discussion on ferrimagnetic (FiM) and
ferromagnetic (FM) case. The Hamiltonian of the magnet coupled with the microwave inside a cavity though its magnetic field component can be written as $\mathcal{H}=\mathcal{H}_{\mathrm{FiM}} + \mathcal{H}_{\mathrm{ph}} + \mathcal{H}_{\mathrm{int}}$, where $\mathcal{H}_{\mathrm{FiM}}$, $\mathcal{H}_{\mathrm{ph}}$, $\mathcal{H}_{\mathrm{int}}$ are respectively the Hamiltonian for FiM, photon and their interaction that read,
\begin{align}
&\mathcal{H}_{\mathrm{FiM}}=J \sum_{l,\delta}  \mathbf{S}_l \cdot \mathbf{S}_{l+\delta}
- \sum_l (\mathbf{H} - \mathbf{H}_{\mathrm{an}}) \cdot \mathbf{S}_{l},\nonumber\\
&\mathcal{H}_{\mathrm{ph}}=\frac{1}{2}\int \left ( \epsilon_0 \mathbf{E}^2
+ \frac{1}{\mu_0}\mathbf{B}^2 \right ) \mathrm{d}x\mathrm{d}y\mathrm{d}z,\nonumber\\
&\mathcal{H}_{\mathrm{int}}=-\sum_{l} \mathbf{S}_l\cdot \mathbf{H}_f,
\label{fmh}
\end{align}
where $J>0$ is the exchange constant, $\mathbf{S}_l$ is the spin on site $l$, and $\delta$ is the displacement of two nearest spins. $\mathbf{H}$ and $\mathbf{H}_{\mathrm{an}}$ are the external static and anisotropy fields, respectively. $\mathbf{E}$ and $\mathbf{B}$ are electric field and magnetic inductance components of the electromagnetic (EM) wave, while $\mathbf{H}_f$ is the corresponding magnetic field, and $\epsilon_0$, $\mu_0$ are respectively vacuum permittivity and susceptibility.

Following the standard quantization procedures for magnons and photons~\cite{Holstein1940,DFWalls,noteH}, the Hamiltonian can be reformulated as,
\begin{align}
\mathcal{H}&=  \omega_a a^\dagger a + \omega_b b^\dagger b +g_{ab}(a^\dagger
b^\dagger + a b) \nonumber\\
&\ \ \ \  + \omega_c c^\dagger c + g_{ac} \left ( a^ \dagger c^\dagger + a c \right)
+ g_{bc} \left (  b^\dagger c + b c^\dagger \right) ,
\label{Fimh}
\end{align}
in which, $\omega_a=H_{\mathrm{ex,b}} + H_{\mathrm{an,a}}+H, \,\omega_b=H_{\mathrm{ex,a}} + H_{\mathrm{an,b}} - H$, $H_{\mathrm{ex,\mu}}=2zJS_\mu^2$~ ($\mu=a,b$), $S_\mu$ is the magnitude of spin vector, $z$ is the coordination number, $g_{\mathrm{ab}}=\sqrt{H_{\mathrm{ex,a}}H_{\mathrm{ex,b}}}$ is the coupling of magnon modes excited on the two sublattices, $\omega_c$ is the photon frequency, $g_{\mu c}$ ($\mu=a,b$) is the interaction strength between magnons and photons. Here $a^\dagger, b^\dagger, c^\dagger$, $a, b, c$ refer to the creaiton/annhilation operators for magnons and photons, respectively.

The Hamiltonian~(\ref{Fimh}) leads to the quantum Langevin equations,
\begin{align}
&\frac{da}{dt}=-(\gamma_a + i\omega_a)a - ig_{ab} b^\dagger -ig_{ac} c^\dagger +\sqrt{2\gamma_a}a_{\mathrm{in}},\nonumber\\
&\frac{db}{dt}=-(\gamma_b + i\omega_b)b - ig_{ab} a^\dagger -ig_{bc} c +\sqrt{2\gamma_b}b_{\mathrm{in}},\nonumber\\
&\frac{dc}{dt}=-(\gamma_c + i\omega_c)c - ig_{ac} a^\dagger -ig_{bc} b +\sqrt{2\gamma_c}c_{\mathrm{in}},
\label{lang}
\end{align}
where we have introduced the dissipation ($\gamma_a,~\gamma_b,~\gamma_c$) and input quantum noise ($a_{\mathrm{in}},b_{\mathrm{in}},c_{\mathrm{in}}$) for each mode according to the standard fluctuation-dissipation theorem~\cite{DFWalls}. Here the noise has zero mean average and delta-correlation in the form of $\langle a_{\mathrm{in}} (t) a_{\mathrm{in}}^\dagger (t')\rangle = \langle b_{\mathrm{in}} (t) b_{\mathrm{in}}^\dagger (t')\rangle = \langle c_{\mathrm{in}} (t) c_{\mathrm{in}}^\dagger (t')\rangle =  \delta(t-t')$ \cite{notenth}.

To quantify the entanglement of arbitrary two modes in the system, we first define the quadrature operators of each mode as $X_\mu=(\mu+\mu^\dagger)/\sqrt{2},~ Y_\mu=(\mu-\mu^\dagger)/\sqrt{2}i$ ($\mu=a,b,c$). Then the Langevin equation~(\ref{lang}) can be reformulated in a matrix form, $d\mathbf{u}/dt=\mathbf{M} \cdot \mathbf{u}+\Lambda \cdot \mathbf{u}_{in}$, where $\mathbf{u}=(X_a,Y_a,X_b,Y_b,X_c,Y_c)^T$, the noise $\mathbf{u_{\mathrm{in}}}=(X_{\mathrm{in,a}},Y_{\mathrm{in,a}},X_{\mathrm{in,b}},Y_{\mathrm{in,b}},X_{\mathrm{in,c}},Y_{\mathrm{in,c}})^T$, the matrix $\Lambda=diag(\sqrt{2\gamma_a},\sqrt{2\gamma_a},\sqrt{2\gamma_b},\sqrt{2\gamma_b},\sqrt{2\gamma_c},\sqrt{2\gamma_c})$,
and the drift matrix,
\begin{equation}
\mathbf{M}=\left (
\begin{array}{cccccc}
  -\gamma_a & \omega_a & 0 & -g_{\mathrm{ab}} & 0 & -g_{\mathrm{ac}} \\
  -\omega_a & -\gamma_a & -g_{\mathrm{ab}} & 0 & -g_{\mathrm{ac}} & 0 \\
  0 & -g_{\mathrm{ab}} & -\gamma_b & \omega_b & 0 & g_{\mathrm{bc}} \\
  -g_{\mathrm{ab}} & 0 & -\omega_b & -\gamma_b & -g_{\mathrm{bc}} & 0 \\
  0 & -g_{\mathrm{ac}} & 0 & g_{\mathrm{bc}} & -\gamma_c & \omega_c \\
  -g_{\mathrm{ac}} & 0 & -g_{\mathrm{bc}} & 0 & -\omega_c & -\gamma_c
\end{array} \right ).
\label{Mxyz}
\end{equation}
According to the Routh-Hurwitz criterion, the system is stable if all the eigenvalues of $\mathbf{M}$ has negative real parts~\cite{Dejusus1987}, which is almost automatically satisfied for an AFM that will be explained below.

The linearity of the Langevin equation together with the Gaussian nature of the noise imply that the steady state of the three-mode system is a Gaussian state and it is fully characterized by a steady covariance matrix $\mathbf{V}$
defined as $V_{ij}=\langle u_i u_j + u_j u_i \rangle /2$, which can be obtained by solving
the Lyapunov equation~\cite{Vitali2007}, $\mathbf{M}\mathbf{V} + \mathbf{V}\mathbf{M}^T=-\mathbf{D}$, where $\mathbf{D}=diag(\gamma_a,\gamma_a,\gamma_b,\gamma_b,\gamma_c,\gamma_c)$.
The strength of entanglement of two modes of interest is quantified by the logarithmic negativity defined as, $E_N=\max \left[ 0, -\ln[2\eta^-]\right ]$, where $\eta^-= \sqrt{ \sum (\mathbf{V'}) - \left [ \sum \mathbf{V'}^2-4\det \mathbf{V'} \right ]^{1/2}}/\sqrt{2}$, $\sum \mathbf{V}'$ and $\det \mathbf{V}'$ are the two symplectic invariants of the reduced covariance matrix $\mathbf{V}'$ of two modes \cite{Adesso2007,note02,note_regime}.
\begin{figure}
\centering
\includegraphics[width=0.48\textwidth]{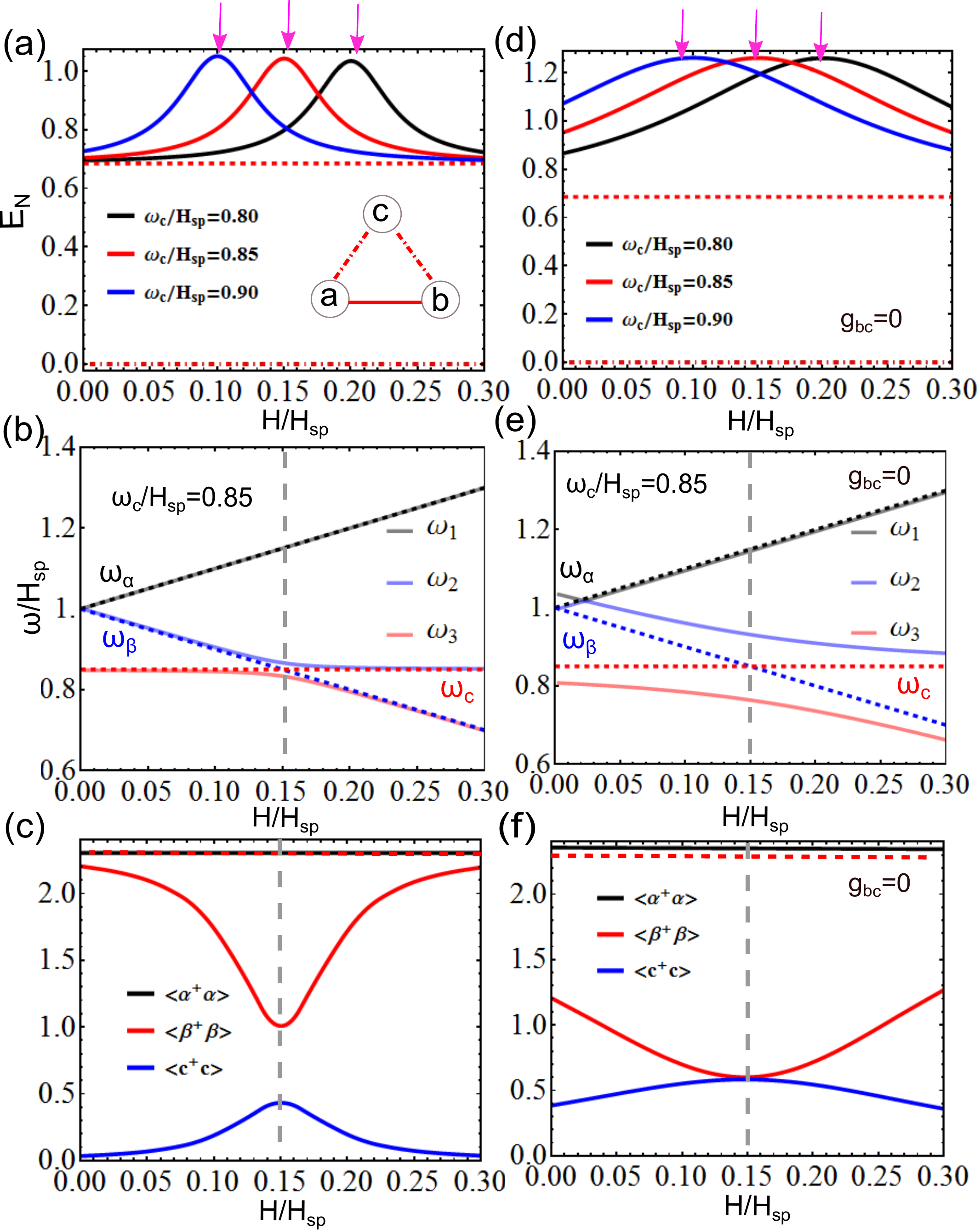}\\
\caption{(a) The enhanced entanglement $E_N$ between two magnons $a$ and $b$ in the presence of light mode $c$ as a function of external field $H/H_{\mathrm{sp}}$, for photon frequency $\omega_c/H_{\mathrm{sp}}=0.80$ (black line), $0.85$ (red line), $0.90$ (blue line), respectively. The horizontal dashed line at $E_N=0.6851$ is the magnon-magnon entanglement without light while the merged dot-dashed lines at $E_N=0$ indicate no entanglement between $a-c$ and $b-c$. (b) The dispersion relations for an antiferromagnet with $\omega _c/H_{\mathrm{sp}}=0.85$ and $g_{ac}=g_{bc}=0.01H_{\mathrm{ex}}$. Obviously, the optical magnon band ($\omega_\alpha$) is left unchanged, while the acoustic band ($\omega_\beta$) is hybridized with photon mode ($\omega_c$) to form modes $\omega_{2,3}$. (c) The population of modes $c$ and $\alpha, \beta$. The red dashed line represents the occupation of $\alpha/\beta$ mode without light. (d-f) are similar to (a-c) but with $g_{ac}=0.01H_{\mathrm{ex}}$ and $g_{bc} =0$. Other parameters are $g_{ab}=H_{\mathrm{ex}}$, $H_{\mathrm{an}}=0.0163H_{\mathrm{ex}}$, $\gamma_\mathrm{m}=0.001H_{\mathrm{ex}}$ and $\gamma_c=0.003H_{\mathrm{ex}}$. For simplicity in the following all the parameters are expressed in units of $H_{\mathrm{ex}}$.}
\label{fig2}
\end{figure}

{\it Antiferromagnetic case.---} In an AFM, the sublattice permutation symmetry implies $S_a=S_b\equiv S,~\gamma_a=\gamma_b \equiv \gamma_m$, $g_{ab}=H_{\mathrm{ex}}\equiv2zJS^2$, $g_{\mathrm{ac}}=g_{\mathrm{bc}}\equiv g_{\mathrm{mp}}$. In this case, we can analytically solve the eigenequation $\det(\lambda \mathbf{I} - \mathbf{M})=0$, and obtain $\lambda_{1,2,3,4}=-\gamma_m - i (H_{\mathrm{sp}}\pm H)$ for $g_{\mathrm{mp}}=0$, where each eigenvalue is double degenerate and $H_{\mathrm{sp}}=\sqrt{H_{\mathrm{an}}(H_{\mathrm{an}}+2H_{\mathrm{ex}})}$ is spin-flop field of the system. Since the dissipation rate $\gamma_m>0$, the system can always reach a stable steady state, regardless of the initial states. This stability condition imposes less constrains on the parameters than those in the traditional three-mode optomechanical systems, where a strict constrain on the coupling strengths and the dissipation rates are required~\cite{Genes2008}. This conclusion is also valid in the case of $g_{\mathrm{mp}} \neq 0$~\cite{note_stable}.

Figure~\ref{fig2}a shows that the magnon-photon entanglement is zero (merged dot-dashed lines at $E_N=0$) while the magnon-magnon entanglement is significantly enhanced compared with the value in the absence of light (dashed line around $\ln 2$~\cite{note_regime}). As the photon frequency is tuned from $\omega_c/H_{\mathrm{sp}}$=0.80 (black), 0.85 (red) to 0.90 (blue), the maximum enhancement locates at $H/H_{\mathrm{sp}}$ = 0.20, 0.15 and 0.10, respectively. To analyze the essential physics, we first diagonalize the antiferromagnetic Hamiltonian $\mathcal{H}_{\mathrm{FiM}}$ in Eq.~(\ref{fmh}) by introducing two Bogoliubov modes $\alpha = \cosh \theta a- \sinh \theta b^\dagger $, $\beta = -\sinh \theta a^\dagger + \cosh \theta b$, where $\tanh 2\theta = -2g_{ab}/(\omega_a + \omega_b)$. In terms of the Bogoliubov modes, the Hamiltonian~(\ref{Fimh}) becomes
\begin{align}
\mathcal{H} = ~&\omega_\alpha \alpha^\dagger \alpha + \omega_\beta \beta^\dagger \beta + \omega_c c^\dagger c \nonumber \\
&\ \ + g_{\alpha c}(\alpha^\dagger c^\dagger+ \alpha c )+ g_{\beta c}( \beta c^\dagger+ \beta^\dagger c ),
\label{abeta}
\end{align}
where $g_{\alpha c} = g_{ac} \cosh \theta + g_{bc} \sinh \theta,~g_{\beta c} = g_{ac} \sinh \theta + g_{bc} \cosh \theta$, and $\omega_{\alpha,\beta} = \pm H + H_{\mathrm{sp}}$ represent the optical and acoustic magnon bands, respectively.

Let us first analyze the position of maximum enhancement of magnon-magnon entanglement in the presence of photons. The two eigenmodes of magnons ($\omega_{\alpha,\beta}$) together with the photon mode for $\omega_c/H_{sp}=0.85$ are plotted in Fig.~\ref{fig2}b as dashed lines. Now we immediately see that the maximum enhancement of entanglement, as seen in Fig. \ref{fig2}a, appears when the strong coupling occurs by adjusting the external field $H/H_{\mathrm{sp}} =0.15$ so that the photon frequency becomes resonant with the frequency of the acoustic magnon $\beta$, i.e., $\omega_\beta= \omega_c$. Due to the coupling of magnons with photons, a gap should open near the crossing point $H/H_{\mathrm{sp}}=0.15$ as shown by solid blue and red lines in Fig.~\ref{fig2}b, i.e., the eigenmodes $\omega_{1,2,3}$ of the coupled system become superpositions of the magnon modes and photon mode. Here acoustic magnon frequency $\omega_\beta$ and photon frequency $\omega_c$ superpose to form the eigenfrequencies $\omega_{2}$ and $\omega_{3}$, while the optical magnon mode $\omega_{\alpha}$ is left unchanged as a dark mode~\cite{Yuan2017apl}, indicated by $\omega_1=\omega_\alpha$ (merged black solid and dashed lines in Fig. \ref{fig2}b). The depth of anticrossing gap at $H/H_{\mathrm{sp}} =0.15$ indicates the effective coupling strength between magnons and photons.

We now proceed to explain the enhancement of the magnon-magnon entanglement in the presence of light mode. In terms of the two-mode squeezing operator $S(\theta) = \exp [\theta (ab-a^\dagger b^\dagger)]$~\cite{Braun2005}, the magnon eigenmodes can be reformulated as $\alpha= S(\theta) a S^\dagger (\theta)$, $\beta= S(\theta) b S^\dagger (\theta)$, and thus the joint ground state of $\alpha$ and $\beta$ modes is a two-mode squeezed state $|\theta \rangle = S(\theta) | 0_a, 0_b \rangle$ with entanglement $2 |\theta|$~\cite{note_squeeze}, where $|0_a,0_b \rangle$ is the joint vacuum of magnon modes $a$ and $b$. According to Eq. (\ref{abeta}), the photon mode couples with the acoustic magnon mode $\beta$ by a beam-splitter-type interaction and thus serves as a cooling bath to cool the mode $\beta$ toward its ground state~\cite{YDWang2013}. Note that the ground state entanglement $E_N = 2|\theta| \approx \mathrm{arctanh} (1+H_{\mathrm{an}}/H_{\mathrm{ex}})^{-1}\approx 2.41$ is larger than the entanglement in the absence of the light ($\ln2$). This cooling mechanism consequently enhances the entanglement of two magnons. As the other Bogoliubov mode $\alpha$ is a dark mode that is non-resonant with the photons, its role in the cooling process is thus neglectable. To demonstrate this point, we plot the average population of $\alpha$, $\beta$ and $c$ in Fig.~\ref{fig2}c as black, red, and blue lines, respectively. Clearly, the occupancy of mode $\beta$ is reduced towards its ground state under cavity cooling and takes on minimum while the population of photon mode $c$ takes on a maximum near the anticrossing point, where the strong coupling between $\beta$ and $c$ occurs and the cooling effect is most efficient. The population of dark mode $\alpha$ is almost unchanged when we tune the field, as it is decoupled from the cavity mode $c$.
This means that the achieved state is not an ideal two-mode squeezed vacuum state as $\alpha$ is not cooled, but Fig.~\ref{fig2}a shows that the entanglement is significantly enhanced $E_N\approx 1.05$.

To further testify the physics of cavity cooling, we artificially tune $g_{bc}=0$ and find that the magnon-magnon enhancement becomes even stronger, as shown in Fig.~\ref{fig2}d. This is because the coupling strength $|g_{\beta c}|$ between modes $\beta$ and $c$ becomes larger when $g_{bc} =0$ compared with the case of $g_{bc}\neq 0$~\cite{coupling} (evidenced by the larger frequency split of the two anticrossing modes at the resonance shown in Fig.~\ref{fig2}e) and thus induces more efficient cooling of magnons (indicated by the lower population of mode $\beta$ in Fig.~\ref{fig2}f) as well as stronger enhancement of magnon-magnon entanglement (Fig.~\ref{fig2}d). Similar enhancement is observed when $g_{ac}=0$.
\begin{figure}
  \centering
  \includegraphics[width=0.48\textwidth]{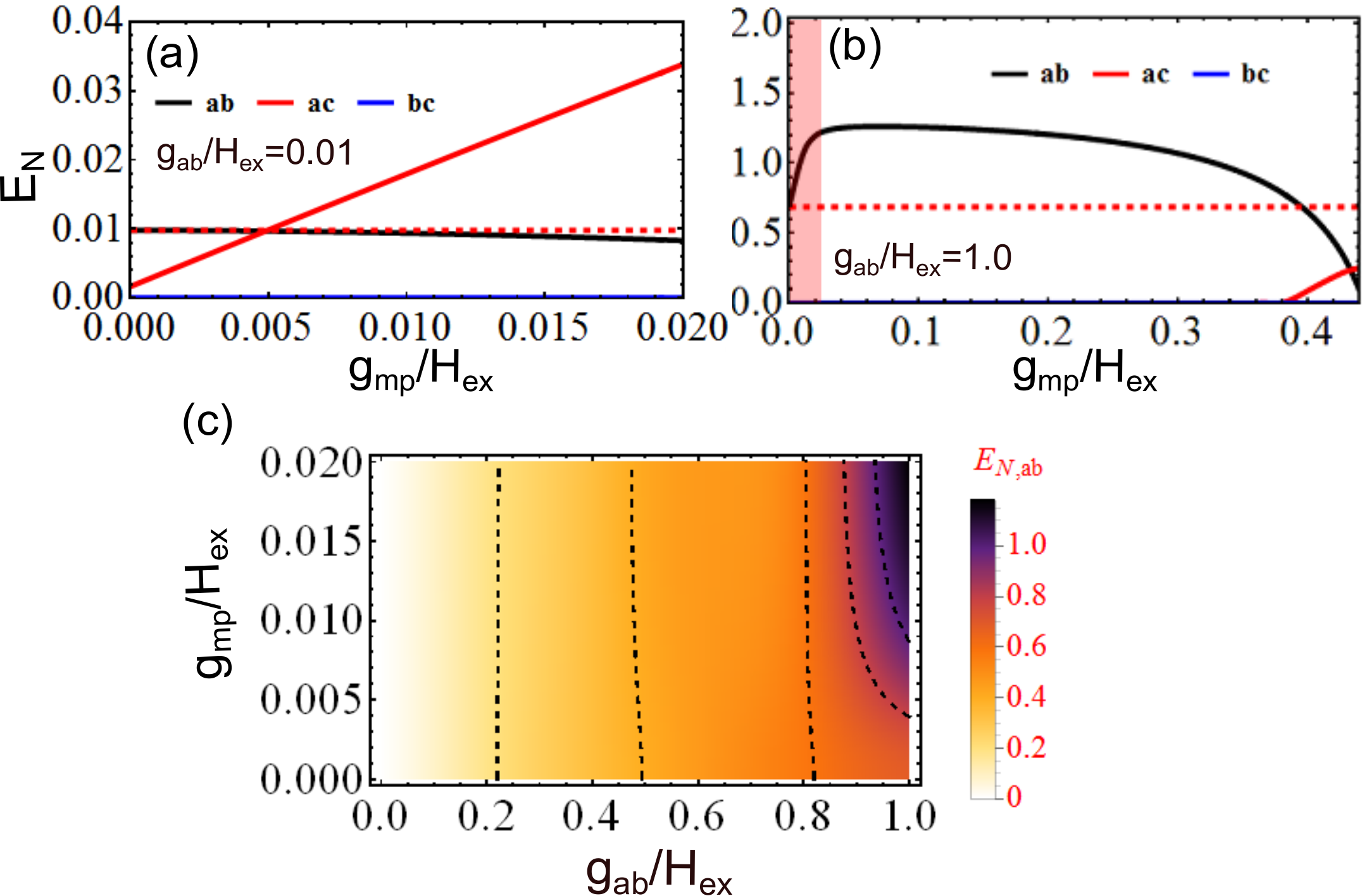}
  \caption{The entanglement as a function of magnon-photon coupling strength $g_{mp}$
  when the system is in the strong coupling regime $g_{\mathrm{ab}}/H_{\mathrm{ex}}=0.01$ (a) and in the deep strong coupling regime $g_{\mathrm{ab}}/H_{\mathrm{ex}}=1$ (b), respectively. The red dashed lines represent the magnon-magnon entanglement in the absence of the photon mode. (c) The steady magnon-magnon entanglement distribution in the $g_{\mathrm{ab}}-g_{\mathrm{mp}}$ phase plane. The right side marks the deep strong coupling regime, where a considerable enhancement is observed. Other parameters are $H/H_{\mathrm{sp}}=0.15, \omega_c/H_{\mathrm{sp}}=0.85$, $\gamma_\mathrm{m}=0.001H_{\mathrm{ex}}$, $\gamma_c=0.003 H_{\mathrm{ex}}$.}
\label{fig3}
\end{figure}

We point out that the enhancement of the magnon-magnon entanglement  appears only in the deep strong coupling regime ($g_{ab} \sim \omega_{a,b}$), and it is absent in the weak and even in the strong coupling regimes where $g_{ab} \sim 0.01-0.1\omega_{a,b}$. This is clearly seen from Fig.~\ref{fig3}a that in the strong coupling regime, $g_{\mathrm{ab}}/H_{\mathrm{ex}}=0.01$, the entanglement between $a$ and $c$ starts to appear when $g_{\mathrm{mp}}$ increases from its $g_{\mathrm{mp}}=0$ value, whereas the entanglement between $a$ and $b$ slightly decreases. This is expected since a part of entanglement between $a$ and $b$ is transferred to mode $c$ through the beam-splitter-type coupling between modes $b$ and $c$. On the other hand, in the case of a deep strong coupling regime illustrated in Fig.~\ref{fig3}b for $g_{ab}/H_{\mathrm{ex}}=1.0$, the significant enhancement of the magnon-magnon entanglement appears and it remains large when $g_{\mathrm{mp}}$ is increased, whereas the entanglement between $a-c$ is completely suppressed. When the coupling strength $g_{\mathrm{mp}}$ is further increased such that it approaches the value of $g_{ab}$, the competition starts and the entanglement between $a-c$ appears as well, as indicated in Fig.~\ref{fig3}b. A complete phase diagram of these two cases is shown in Fig. \ref{fig3}c. The absence of the enhancement of the magnon-magnon entanglement in the weak and strong coupling regimes can be well understood in the cavity cooling scheme. In the strong coupling regime, the entanglement between $a-b$ without the light mode is given by $E_N \approx \ln(1+g_{ab}/(H_{\mathrm{ex}}+H_{\mathrm{an}})) \approx g_{ab}/(H_{\mathrm{ex}}+H_{\mathrm{an}})$, which is almost equal to the entanglement of the joint vacuum
$|2\theta| \approx |\tanh 2\theta| = g_{ab}/(H_{\mathrm{ex}}+H_{\mathrm{an}})$ since $|2\theta| \ll 1$ \cite{note_regime}. Hence, the cooling of magnon mode to its vacuum does not induce considerable enhancement of the magnon-magnon entanglement, while the competition of creating entanglement between $a$ and $c$ reduces the original entanglement between $a$ and $b$.

{\it Ferri-/Ferro-magnetic case.---} For a two-sublattice ferrimagnet ($S_a \neq S_b$,
$g_{ac} \neq g_{bc}$), the enhancement of magnon-magnon entanglement is similar to the antiferromagnet case once the deep strong coupling condition is satisfied, i.e., $g_{ab} \sim \omega_a, \omega_b$. A ferromagnet corresponds to the limiting case of $S_a \rightarrow 0$ in which $g_{ab} \propto S_aS_b\rightarrow 0$, and then the physics becomes different. Specifically, for a single sublattice ferromagnet coupled with the light mode, Eq.~(\ref{Fimh}) is reduced to $\mathcal{H}=\omega_b b^\dagger b + \omega_c c^\dagger c +g_{bc}(b^\dagger c + bc^\dagger)$, the spectrum of which takes the form, $\omega_{1,2}= \frac{1}{2}\left [ (\omega_b + \omega_c) \pm \sqrt{(\omega_b - \omega_c)^2 + g_{bc}^2}\right ]$, which has an anticrossing near the point $\omega_b=\omega_c$ \cite{note04}. However, the stable state is not an entangled state, because the magnon mode $b$ is coupled to the photon $c$ by a beam-splitter-type interaction. Note also that the parametric-down-conversion-type coupling of $a$ and $c$ in AFM case does not lead
 to an entanglement state, either. These findings clearly show that the magnon-photon coupling is not a sufficient condition for magnon-photon entanglement.

{\it Conclusions.---} In summary, we have studied the entanglement properties of magnons and photons inside a cavity and find that the entanglement between magnons and photons are very weak. Instead, the magnons excited on the two sublattices of an AFM are strongly entangled and this entanglement can be enhanced to its maximum when the magnons are coupled to the photon in the resonant condition. The maximum enhancement increases monotonically with the coupling strength between magnons and photons while it optimizes at a particular dissipation rate of the cavity~\cite{note_dissipation}. We ascertain that such enhancement comes from the cavity cooling effect and it is a unique feature of the antiferromagnet with the deep strong coupling between two magnons and it disappears for normal system with strong coupling. To measure the entanglement enhancement, one may couple an AFM insulator such as $\mathrm{MnF_2}$ and $\mathrm{NiO}$ with high-quality cavity/co-planar waveguide and measure the noise level of the system.

HYY acknowledges the financial support from National Natural Science Foundation of China (NSFC) Grant No. 61704071. QH acknowledges NSFC Grants No. 11622428 and No. 61675007, the National Key R\&D Program of China (2016YFA0301302 and 2018YFB1107200), MHY acknowledges support by Natural Science Foundation of Guangdong Province (2017B030308003), Guangdong Innovative and Entrepreneurial Research Team Program (2016ZT06D348), and Science, Technology and Innovation Commission of Shenzhen Municipality (ZDSYS20170303165926217 and JCYJ20170412152620376).

\clearpage
\onecolumngrid


\begin{thebibliography}{}
\bibitem{notenth} In general, $\langle c_{\mathrm{in}} (t) c_{\mathrm{in}}^\dagger (t')\rangle = (n_{\mathrm{th}}+1) \delta(t-t')$, where $n_{\mathrm{th}}$ is the Bose-Einstein distribution function for the photons.
    Here we consider the low temperature case such that $n_{\mathrm{th}} \ll 1$ and we choose $n_{\mathrm{th}}=0$.
\bibitem{note04} At this point, we can solve the eigenequation
$det|\lambda \mathbf{I}-\mathbf{M}| =0$ and obtain
$\lambda_{1234}=-\gamma \pm i (\omega_c\pm g_{bc})$ where we have assumed
$\gamma_b=\gamma_c=\gamma$ for simplicity. This implies that the system can always reach a stable state.
\end{thebibliography}
\end{document}